# Life, Intelligence and Multiverse

Zoltán Galántai
Budapest University of Technology and Economics (Hungary)

*"Perhaps ultimate questions are not questions at all."*
*Roger Scruton*

**Abstract**
Hypothetical existence of other universes gives an opportunity not only to extend the scope of physics, but the scope of biology, SETI, and METI as well. Some steps of the development of alien life concept shall be briefly summarized, then the multiverse proposal shall be used as a framework of interpretation to introduce an extended taxonomy of possible or at least imaginable types of life and intelligence based on either different biochemistry or physics. Some consequences shall be presented about SETI and METI in connection with both multiverse hypothesis and anthropic principle.

**From natural history to astrobiology to xenolife**
*"The biologist passes, the frog remains." (Jean Rostand)*

Although both the search for alien intelligences and the multiverse hypothesis are relatively new, they have some historical roots.
Our ancestors believed in non-human intelligences including gods and goddesses, angels, spirits and monsters who shared the earth with them, [Michaud 2007, 9] but the idea of world creation by a god (demiurge) in philosophy appeared for the first time only in Plato's dialogue entitled Timaeus. [Rubinstein 2014, 33]
The world system of the Middle Ages was hierarchical. It was based on the idea of the "great chain of beings" which was inherited from the classical period. This included a belief in the existence of both superior and inferior (non-spiritual) creatures who lived with us. [Case-Winters 2007, 149]
After the invention of the telescope the other planets of the Solar System were regarded to be similar to the earth, therefore it was believed that they were populated by other "humans", animals and plants. The first literary work inspired by the Copernican system advocating extraterrestrial life on the moon was Kepler's Somnium in the early 17[th] century, and it depicted both the curious animals of the moon and a lunar society, [Basalla 2006, 25] but the interest was mainly focused on the "intelligent" life forms and the thinkers didn't play close attention to the "inferior" beings at that time.
In part because biology is a newcomer compared either to physics or astronomy. Its field was divided into the realms of natural history and medicine for centuries and the connections between these parts was weak even in the 18[th] century. Lamarck was one of the firsts to use "biology" as a term around 1800, but this biology neither contained taxonomy nor natural history then and it annexed every field of the life sciences only in the 19[th] century. [Junker 2004, 7 - 8]
It is a more recent notion than a complex and extended biology that other life forms' biochemistry can be different from ours. For example, H. G. Wells played with the idea of a "silicon – aluminum organism" at the end of the 19[th] century anticipating the modern SF's "silicon based life". [Darling 2002, 14]
Not everybody accepted this approach. An American biologist named John Pratt already in 1883 argued defending a carbon chauvinist standpoint that life was unimaginable without heat, light, oxygen, carbon, etc., namely: An alien life was necessarily similar biochemically to the life on earth. But his critic, Charles Morris claimed soon that the appearance of life was not only possible, but was probable without carbon on other planets with different environments. [Crowe 1986, 466] It was a beginning of the debate about the "chemical necessities" of life, that is, whether life is necessarily based on certain chemical element or not.

The threat of back contamination by astronauts or returning samples to earth required to establish a new science at the beginning of the Space Age. [Dick 1996, p. 138] It was originally called to be "extraterrestrial biology" or "exobiology"; the NASA began to use the term astrobiology only in 1996. [Dick 2013, p. 140] It is not a surprise that this beginning with a wish for planetary protection more or less determined astrobiology for carbon chauvinism although the definition of this new field remained slightly vague for decades.

On the other hand, Morris' approach was extended into a new interpretation called xenoscience by Jack Cohen and Ian Stewart around 2000. This approach is positioned as an antithesis of the traditional school of astrobiology. According to it, "astrobiology is the science of earthlike planets supporting earthlike life", while xenoscience is "to focus on the *biology* of aliens." Its central concept is that although so much is known about earthly organism, it is a mistake to think that life as a general phenomenon is known. [Cohen and Stewart 2002, 5 – 6]

Taking into consideration the differences between possible life forms, one can introduce the term of the "xenolife" as a tool to distinguish creatures which are based on earthly biochemistry form those which are based on different chemical combinations (silicon, non-water solvents, etc.). After all, it is a fundamental question whether the life in the universe can be described by astrobiology or by xenolife; and whether our universe is dominated by biochemistry similar to ours or several biochemistries exist.

**Interlude: multiverse and habitability**
*"The great tragedy of science – the slaying of a beautiful hypothesis by an ugly fact". (Thomas Huxley)*

Before examining some possible interpretations of xenolife, it is worth mentioning the multiverse proposal, because it can influence both the search for the alien life and either SETI (search for extraterrestrial intelligence) or METI (messaging to them).

According to British astronomer Bernard Carr, the history of cosmology can be divided into five periods.

1. The first was the geocentric world view: The presence of life was restricted to the earth and its spheres which were inhabited by angels and spirits.

2. It was followed by the heliocentric era after the publication of Copernicus's book in 1543. The Solar System was believed to be the whole universe and besides moon "selenites" and other planet dwellers, even inhabited comets appeared in Fontenelle's book in the late 17$^{th}$ century [Fontenelle 1803, 122]; and even the surface of the Sun was considered to be habitable by William Herschel a hundred years later. [Kavaler and Veverka 1981, 48].

3. The Milky Way was interpreted as disc of stars from the mid-18$^{th}$ century and it was regarded to be the whole universe. This galactocentric view dominated the human thought even in the year of the publication of Einstein's General Relativity in the early 20$^{th}$ century.

4. Although it had some forerunners (e.g. Kant), the cosmocentric view began to rule the astronomy only after Hubble's announcement the discovery of other galaxies in the mid-1920's. The beginning of the cosmocentric era in the first part of the 20$^{th}$ century coincided with the popularity of James Jeans' hypothesis about the exceptionality of the planetary system formation. It suggested the rarity of life in the universe, but the belief in the earth's exceptionality weakened around 1950 [Dick 1996, 180] allowing the acceptance of the astrobiological thought–although nothing supported the existence of exoplanets at that time. But it is still not sure if the existence of other, even "earthlike" planets are a guarantee for extraterrestrial life, since a planet's similarity to our home is only a necessary but not a sufficient condition of the earthlike evolution.

5. The last phase in Carr's classification is the multiverse concept. This seems to be more or less acceptable for cosmologists, since "The longer we have studied the Universe, the larger it has become," says Carr [2007, 8-9]. This space scale extension is not necessarily an everlasting process, but introducing the multiverse idea gives an opportunity to study habitability in a new context.

A multiverse is an "ensemble of universes or expanding domains like the one we see around us" [Ellis 2007, p. 387] and there are four possible multiverse concept variations. The separation of a universe from the others can be based either on space (these are spatially multiple universes); on time (if they are temporally multiplied); or on other dimensions. This third kind was already proposed by Leibniz in 1686. [Gale 1998, p. 196 – 199] The fourth version of multiverse models was introduced by John D. Barrow in 1992. He imagined a "network of mathematical possibilities" or "mathematical structures that correspond to the self-conscious beings" [Barrow 1992, 282]. As it was phrased by Barrow, this system is "an ensemble of universes in which all the possible axiomatic systems of mathematics holding." This approach was improved by Max Tegmark and it is related to modal realism stating that every possible world is as real as our actual one. It consists every mathematically possible universes' copies as physically existing objects in infinite number. Most universe theorists accept one or two models, but not all of them, and Tegmark's proposal seems to be the most exotic. [Rubinstein 2014, 17]

The main problem with the multiverse proposal is that "only one universe is to be observed, and that we effectively can only observe it from one space–time point," says Ellis. [1998, p. 119] But the existence of the multiverse cannot be excluded today and so it is theoretically possible that some other universes are habitable.

**Gigatrajectories, life and intelligence**
*"It has yet to be proven that intelligence has any survival value." (Clarke's Law of Evolution)*

Unless one accepts the creation either by the religions mighty Creator or by a quantum fluctuation which results Boltzmann-brains, [Albrech – Sorbo 2008, 5] a hypothesis about life, intelligence and their relations has to be based on the assumption that life is an inevitable precondition for intelligence–at least at its beginnings. Even the so-called postbiological intelligence hypothesis is only a description of a transition from intelligent biological to intelligent non-biological forms. It is based on a presumption that evolution can lead intelligence which produces culture and culture's evolution leads necessarily to the birth of non-biological, intelligent entities, since it offers longevity which is desirable for every being. [Dick 2006, 2]

According to a taxonomy of evolutionary milestones, the emergence of life was the first so-called megatrajectory; it was followed by the prokaryote diversification; then eukaryotic cells appeared; then the rise of the multicellular life forms occured; the "invasion of the land" was the $5^{th}$ and the rise of intelligence was the $6^{th}$ megatrajectory. The first four megatrajectories were based on increasing inner (genomic, morphological, etc.) complexities. The appearance of life on the lands caused an ecological expansion, and the rise of human intelligence gave an opportunity for the invasion of every possible environment. [Knoll and Bambach 2000, 2 - 9] This invasion perhaps covers the outer space as well.

The post biological intelligence can be interpreted as the $7^{th}$ megatrajectory [Cirkovic, Dragicevic and Beric-Bjedov 2005, 96]. It is nothing more in a certain sense than expressing a belief in the power of intelligence that can modify not only the environment, but itself, too. It would be exciting to know what would be the next megatrajectory–providing, but not necessarily permitting the existence of an $8^{th}$ phase.

This taxonomy is focuses at least partly on parochial details of the earthly evolution, since its main aim is to give a description of the life's history on earth: For example, the fifth megatrajectory would never occur on a water covered planet.

It is possible to reinterpret this classification focusing on those factors that are fundamental for the whole history of the universe where, According to Freeman Dyson, three kind of phenomena can occur: "normal physical processes"; "biological processes" and "communication between life forms existing in different parts of the universe" [Dyson 1979, 1]. These can be interpreted as "gigatrajectiores," because opposite to certain megatrajectories, these necessarily appear in any universe which is populated with intelligent observers. According to this model, the first gigatrajectory is characterized by the domination of lifeless matter; the appearance of life is the second gigatrajectory and finally the appearance of intelligence becomes dominant.

Accepting the possibility of universes determined by different circumstances and/or laws, one can examine the relations between the second (life) gigatrajectory and the third one (intelligence).

For example, although our universe's conditions are appropriate for intelligent life, the possibility of other biofil, but not intelligence friendly universes seem to be possible and perhaps there are universes where the progress ended with the first gigatrajectory.

The presence of the second gigatrajectory is a precondition of the third, but intelligence can emerge as a result of different combinations of possibilities and/or necessities. (As it is shown in table 1.)

*Table 1. Possible relations between life and intelligence*

| type | life | intelligence | in our universe? |
|---|---|---|---|
| T1 | impossible | impossible | no |
| T2 | possible | impossible | no |
| T3 | possible | possible | perhaps |
| T4 | possible | inevitable | perhaps |
| T5 | inevitable | impossible | no |
| T6 | inevitable | possible | perhaps |
| T7 | inevitable | inevitable | perhaps |

One can imagine a universe where life is possible, and after its rise, then the appearance of intelligence is inevitable (T4); or a universe where both life and intelligence is a necessity (T7) etc.

This taxonomy partly overlaps the domain of the Anthropic Principle and offers some new variations.

The Weak Anthropic Principle (WAP) means that „our location in the Universe is necessarily privileged to the extent of being compatible with our existence as observers," since an intelligent being without the special parameters of fundamental physical constants would not be able to come into existence. The SAP states that this fine-tuning is not an accident, and the universe necessarily able to „admit the creation of observers within it at some stage." [Barrow and Tipler, 1986, 1 - 2]. In other words: SAP's supporters believe in the inevitability of the appearance of both life and intelligence in our universe (T7), so a T7 world can be defined as "fine-tuned" for SAP.

Besides the question whether there are other, either biofil or lifeless universes, it is not known if our universe can be categorized as T7 or it belongs to T3 (possible life and possible intelligence) or to T6 (inevitable life and possible intelligence), etc.

**Variations for the chemistry of life**
*"Living organisms are created by chemistry. We are huge packages of chemicals." (David Christian)*

It is easy to find similarities between Pratt's standpoint and some versions of the Anthropic Principle. Carbon chauvinism claims that there is only one life-friendly combination in chemistry, while the AP claims that the physics allows one and only one life-friendly combination. However, Anthony Aguirre pointed out that there was a class of cosmologies with totally different parameters where both the carbon and the heavy elements could form. So the presence of different parameters don't necessarily excludes the appearance of intelligent life [Aguirre 2001].

Following an analogous reasoning, Cohen and Stewart argue that there is a "phase-space" of possible biofil universes, and although a change of a single constant's parameter would result a lifeless world, it does not mean necessarily the impossibility of other, life-friendly combinations [Cohen and Stewart 2002, 18]. Similarly, Alesandro Jenkins and Gilad Perez examined the consequences of the change of light quarks' masses, and according to them, some "alternative universes" with different life-friendly chemistry is possible [Jenkins and Perez 2010, 48].

Obviously, the presence of certain chemical elements isn't enough. It is impossible to create an earthly creature without carbon, but it is not clear whether those laws that can produce the elements needed for life, would automatically produce those conditions which have to be fulfilled before the second gigatrajectory would occur.

**Towards a Super Strong Anthropic Principle?**
*"A universe with a God would look quite different from a universe without one. A physics, a biology where there is a God is bound to look different." (Richard Dawkins)*

It is an open question whether life is a "cosmic imperative" and it necessarily emerges in the presence of certain conditions governed by physics-like prescriptive (if A, then B type) laws; or it is governed by permissive laws, similarly to the evolution where a phase-space of the opportunities is given, but the rise of a new species depends on certain circumstances, chances, previous developmental paths, etc.

Christian De Duve, who is a well-known representative of the "cosmic imperative" approach, argues that "chance does not exclude inevitability." He uses the analogy of lottery where improbable to win in a single drawing, but some hundred million attempts are more than enough [De Duve 2007, 7.].

According to the SAP, the rise of life and intelligence is an inevitable necessity. The multiverse hypothesis seems to offer an alternative solution, since if one accepts the existence of (infinitely) many universes with different conditions, then it is not a surprise to find ourselves in a biofil environment. Following this logic, Brandon Carter proposed the "World Ensemble" to explain the so-called "Anthropic Coincidences." [Gale 1998, 200] After all, life would not arise in a hostile environment and the existence of the multiverse explains why our life friendly universe exist, claims Carter.

Both de Duve's and Carter's argumentation rely on chance and a large number of attempts, but it remains unanswered why the rise of life is allowed by natural laws in any universe at all. It can be regarded as either a mere luck or a Super Strong Anthropic Principle (SSAP) can be introduced. The Strong Anthropic Principle's scope is "only" our Universe. The SSAP claims that the existence of a biofil universe is a necessity in the ensemble of the universes. Obviously, it is only a theoretical possibility, since even the existence of the multiverse is unproven.

**Possible levels of biology and the habitable time zones**
*"I try to show the public that chemistry, biology, physics, astrophysics is life. It is not some separate subject that you have to be pulled into a corner to be taught about." (Neil deGrasse Tyson)*

One can play with the idea of an extended xenolife that examines not only biofil biochemistries which are different from the earthly biochemistry, but it includes any possible versions of the biology based on different physical laws, constants, etc. of other universes (See table 2.). This approach puts it into a broader context the possible forms of the extraterrestrial life and can serve as a basis of an extended taxonomy of imaginable versions of the biology. (See Table 2.)

*Table 2. Possible types of biology*

|  | name | the level of difference from earthly biology |
|---|---|---|
| BIOLOGY 1 | biology | no difference |
| BIOLOGY 2 | astrobiology | different morphologies |
| BIOLOGY 3 | xenolife | different chemistry |
| BIOLOGY 4 | extended xenolife | different physics |

**BIOLOGY 1** is the earthly life as it is known. Only its existence is a certainty today.
**BIOLOGY 2** is a more complicated story with uncertainties. Astrobiology can be interpreted as an extension of our biology's scope to extraterrestrial life forms, but it is not clear if it means only a same biochemistry, or it includes other necessary similarities, for example, the same DNA base structure.
**BIOLOGY 3** is xenolife's terrain. It is essentially about the alternative forms of biochemistry up to inorganic plasma life which hypothetically populates the interplanetary space [Tsytovich, Morfill, at all, 2007] and to life forms which can be found in either solid hydrogen absorbing infrared energy; or radiant life forms in interstellar clouds. [Shapiro and Feinberg 1998, 280]
Perhaps it is only an anthropic bias to believe that the earth is the etalon of the habitability, since it seems to be possible that some other worlds' (planets' or moons') conditions more suitable for an earth-type evolution. [Heller, Armstrong 2014, 1]
Similarly, it seems to be possible that either life (BIOLOGY 2) or xenolife (BIOLOGY 3) can appear in a different era of the universe. According to some calculations, the conditions allowed to form the chemistry of life less than a mere 20 million years after the Big Bang. [Loeb 2014, 6] If it is true, then even BIOLOGY 2 can arise in different cosmic environments and epochs than ours.
Ad analogiam, since it is not a certainty that only today's conditions are favorable either for xenolife's some forms or for some intelligent life forms, it can be introduced the idea of the "habitable time zones" for both the early period and the very late universe.
The possible appearance of life in the early universe can be interpreted as an argument against the Anthropic Principle, [Loeb 2014, 6] because it suggests that humans aren't the only candidates to be the purpose of fine-tuning.
Of course, it is not a new idea to examine the temporal distribution of extraterrestrial civilizations. Even the earliest form of the Drake Equation contained an L parameter to determine the average span of a civilization's communicative state. [Michaud 2007, 55] But L is implicitly suggests that the same environment is favorable for every intelligent being.
**BIOLOGY 4** is about the life in other universes with different physics: E.g. with different parameters of physical constants or different physical forces. Naturally, it cannot be excluded theoretically that even if there are infinitely many other words, natural laws limit the possibilities of life for BIOLOGY 1 or BIOLOGY2 everywhere. The existence of infinitely many universes does not mean that everything which can be imagined is realized: Only those can be realized that are possible. Therefore it is a question whether besides the earthly biology, either BIOLOGY 3 or BIOLOGY 4 possible.

Two different levels of biology can be distinguished at this point. The evolutionary processes can be observed on earth are universal in the sense that every imaginable life forms are subject of them either in Lake Vostok or on Mars or in another universe. But it is controversial whether evolution's known manifestations are universal or parochial. For example, it is not known whether only carbon can serve as a basis of life (as it was discussed by Wells and others already in the 19th centruy). Of course, if carbon chauvinism is proves to be true then the set of the creatures that belongs to xenolife (BIOLOGY 3) is empty.

**BIOLOGY 4** raises another problems. If there are life forms based on different physics, it either can or cannot lead to the rise of intelligence; or can make it inevitable (as it was shown in Table 1). To make things more complicated, the connections between the physics of other universes and the different biochemistries aren't clear. Even if it is a certainty that our universe permits only the existence of carbon based life, it remains a question whether a different physics allows the appearance of different biofil biochemistries simultaneously in another universe.

The first version of Lee Smolin's "cosmological natural selection" was published in 1992. It is a good example for the difficulties of interpretation. It is based on supposed similarities between biological evolution and a selection mechanism which is based on string theory's results as an answer for the questions raised by the Anthropic Principle. According to Smolin, universes can be born via black holes with slightly different physical parameters, and the more successful variations would produce more baby universes. [Smolin 2006, 1]

Smolin replaces SAP's rigid laws that necessarily produce a biofil universe with a kind of evolutionary process, but it is not verified that this process should exist instead of a law that would prescribe the inheritance of the physical laws/parameters in an unchanged form. So for example it is at least imaginable that every universe's physics is the same as ours. I. e. on the one hand every universe is biofil at a certain epoch of its history and the SSAP can be regarded as a rule which requires it. On the other hand, it is possible that even this scenario would lead to the appearance of life only in a certain period. What is more, the "selection" for black holes and the "selection" for a livable environments are not necessarily the same.

Edward Harrison proposed a modified form of the "natural selection": According to his model, our and infinitely many other Worlds were created by an intelligent being living in another universe. If an artificially created offspring universe fits for intelligence, a new life will evolve and reaching a level of intelligence, this intelligent life form will create new universes. The Result: "Universes unfit for inhabitation… cannot reproduce," but life friendly versions will. So it would explain why the constants fine-tuned for life (and intelligence). [Harrison 1995, 193] But it seems to be not defendable to hypothesize that an intelligence necessarily creates offspring universes.

**Extending the scope of the SETI**

*"We have to wonder, if there is a multiverse, in some other patch of that multiverse are there creatures?" (Janna Levin)*

By the analogy based on the hypothetical relations between biology and xenolife, the scope of the search for extraterrestrial intelligence can be extended to other universes.

If one is willing to accept the existence of infinitely many other universes, then it is inevitable to accept the existence of infinitely many universes which are identical to ours and contain even our identical alter egos, for every possible combinations are realized infinitely many times in this model. [Tegmark 2003, 41] The questions raised by these "copycat cosmoi" are not the same questions that are raised by the search for alien intelligence in multiverse.

Because of the enormous distances only a very strong signal can be detectable from another galaxy, so the modern SETI focused mainly for the search of other civilizations in the Milky Way around 1960, although a Soviet astronomer, named Gennady B. Sholomitsky erroneously identified CTA-102 quasar as a super civilization [Harrison 1997, 233].

To receive a beacon is problematic, but perhaps it is possible to detect macro-engineering objects even from intergalactic distances. This concept is based on Haldane's, Dyson's and those others' proposals who were convinced that macro-engineering would play a key role in both humans' and in an advanced intelligence's space exploration. Notice that this "Dysonian approach to SETI" [Bradbury, Cirkovic and Dvorsky 2011] is not about the real nature of intergalactic civilizations, but only about the human race's observational limitations, although it is a possible tool to find alien civilizations.

This approach assumes that intelligence-driven spatial expansion of the $6^{th}$ megatrajectory happens simultaneously with the increase of the size of the artificial constructions. This can be criticized arguing that it is questionable whether there is an inevitable connection between the level of a civilization's development and their spatial extension, energy consumption, etc. Even if it is a real trend of human development today, it is based only on a snapshot about human civilization's actual stage and it is doubtful whether it can be generalized as a universal law for every civilization. [Galantai 2006].

The distance–detectability problem can be discussed on different spatial levels as well. According to Fred Adams and Greg Laughlin, there are "four important size scales: planets, stars, galaxies, and the universe as a whole." They are our "windows to the universe." [Adams – Laughlin 1999, xiii – xiv] It is not known if all of these windows are inevitable preconditions in a universe where life appears or they are as accidental as the invasion of the lands in the history of life. But the idea of these windows is undoubtedly useful. Applying the logic of Carr's historical sketch about the cosmology to them, the history of SETI can be divided into three eras to date.

1. The first period was the planetary SETI and the search for aliens with optical telescopes in the Solar System from Galileo roughly to Lowell who, trying to interpret the earthly evolution to Mars, was an advocate of the intelligent Martian life around 1900. He was convinced that he was able to observe the Martian canals [Basalla 2006, 75] that were regarded as macroengineering works of a superior intelligence. The spatial extension of the search for alien intelligence is roughly equal to the planetary window in this period and Lowell's research program can be interpreted as a form of Dysonian SETI limited to interplanetary distances.

2. The second era from the mid-$20^{th}$ century was determined by radio telescopes and interstellar distances (i.e. the window of stars). The main aim was to find artificial signs/messages and it taking into account the first, optical period, it can be noticed that the search for radio messages which are sent on purpose represent only a fraction of the history of SETI.

3. The third era is the Dysonian SETI. It corresponds to the window of galaxies and to intergalactic observations. Having been a forerunner of this approach, James Annis studied already in 1999 whether there were detectable signs of the activities of alien civilizations in some nearby galaxies. According to his conclusion, neither our Galaxy, nor M31 or M33 were "transformed into a Type III civilization" [quoted by Carrigan 2010] that were able to possess the energy "of its own galaxy". [Kardashev 1964, 219]

The first two eras of the SETI can be identified by their different size scales and technologies (optical vs radio detection). It seems to be reasonable to suppose that intergalactic SETI would lead to the development of new observational tools, too.

4. One can pair the fourth level (the window of the universe) with another level of SETI. This approach can be identified either as the search for a Type IV civilization that can control the energy of the entire universe [Sagan 1975, 234] or a search for the signs of the existence of a universe creator who acted either as a programmer or a cosmologist (e. g. following a method similar to Smolin's).
5. And perhaps there is a fifth level. The multiverse concept makes possible to extend the search for both life and intelligence to other universes (see Table 3).

*Table 3. Possible levels of SETI*

| level | observational method | scope | Type of SETI |
|---|---|---|---|
| SETI 1 | telescope | within our planetary system | Lowellian |
| SETI 2 | radio telescope | within Milky Way | modern |
| SETI 3 | telescope/radio telescope/other (?) | between galaxies | Dysonian |
| SETI 4 | search for signs of intelligence harnessing extragalactic energy? | the whole Universe | Kardashev Type IV |
| | examining the nature of physical laws? | | created universe |
| SETI 5 | neuron-shining-through-a-wall? other methods? | multiverse | multiverse SETI |

It is worth to make two distinctions in connection with the level of SETI 4. First of all, this hypothetical universe creator is not identical to the God of the religions: he or she is not a divine entity, but only an intelligent being with sufficient knowledge for universe construction.

Second of all, it is not necessarily the same to live in a created universe as to live in a computer simulation where the possibilities are restricted only by the limitations of its creator's imagination or by the laws of logics, but he/she can ignore the laws of physics. Although even a programmer creator is unable to create a not biofil universe that is biofil (since it is a logical contradiction), the life programmed by him or her can be independent from any kind of biochemistry.

Opposite to it, it is not known whether a hypothetical universe creator would be able to construct not "only" a new, independent sub universe, but new physical laws for it, as well. After all, his/her possibilities depend on the physical laws which determine his/her physical world.

Of course, to believe in the existence of a programmer or a cosmologist universe creator is problematic from a methodological point of view. The theists of the 17[th] century argued that a divine and continuous intermeddling was needed to operate the world, but according to the deists of that age, the world operated autonomously as a clockwork in the absence of its Creator. [Brooke 2014, 181]

Both of these opinions were based on the "argument from design", although their conclusions were contrary to each other. Similarly, one can believe either in the existence of a creator or in the existence of a world which was formed solely by natural laws, but it is possible to work out a coherent description about the universe in both cases. So it can be argued that there is no reason to introduce the existence of a world creator, since it adds nothing to the explanation.

As for the SETI 5 level, Robin Collins tries to reverse this logic, arguing from a theistic point of view that the supposed existence of multiverses does not rule out the existence of God. [Collins 2007, 459] Namely, if one believes in the creation of a whole universe, then he/she can accept the existence either a Creator or an intelligent non-divine being who is able for multiverse creation.

This argumentation is based on an unspoken premise that if one can create a universe, then he/she can create a multiverse as well, but it is not a logical necessity, since creation doesn't mean necessarily a creation without any restriction. The existence of a multiverse creator cannot be excluded by pure logic, since it is a matter of fact, but even if someone would be able to ascertain both about the existence of a universe creator and the existence of a multiverse, it doesn't make a necessity the existence of a multiverse creator.

**Multiverse-METI?**
*"The medium is the message." (Marshall McLuhan)*

Moving forward form the SETI 4 level to SETI 5 (see Table 3), it can be propounded as the search for other universes' intelligent habitants and the feasibility of messaging between universes similarly can be examined as a theoretical possibility.

Messaging to extraterrestrial intelligence has a shorter prehistory than of SETI. It was proposed around 1800 to form a huge Pythagorean triangle somewhere on the surface of earth to send optical messages to the Moon or Venus dwellers. [Crowe 1986, 205 – 207] It was undoubtedly a kind of planetary METI based on a kind of macroengineering that was followed by some other optical proposals in the 19th century without any attempt for their realization.

The scene of the first known METI attempt was the Soviet Union in 1962. A radio message ("peace/world, Lenin SSSR") was transmitted from Evpatoria Planetary Radar to the planet Venus. [cplire.ru, n. d.] It was followed by the Arecibo Message and a dozen other radio signal attempts directed to different targets within the Milky Way.

Another, possible way of messaging into galactic distances is building huge artificial objects in the outer space using megascale-engineering technologies. It is a reversal of Dysonian SETI. Jaron Lanier proposed the usage of "gravitational tractors" to rearrange asteroids which would modify the sun's orbit; then our sun' gravitational force would modify other stars' orbits. The result would be a "grapstellation" and an artificial arrangement of a cluster of stars would advertise humans' existence to other dwellers of the Galaxy at the end of the process. [Lanier 2008]

It seems to be tempting to use this method for sending signs over the galactic level by rearranging all the stars' positions of the Milky Way into a huge artificial pattern which can be observable form another galaxies. However, John D. Barrow and Frank Tipler pointed out that it is physically impossible in our universe, since rhis process would take a $10^{22}$ years, but the natural forces would disrupt the pattern within $10^{19}$ years. Perhaps an intelligent race would be able to use bigger "gravitation tractors" than asteroids in the future [Davies 1994, 113 – 114] and it seems to be possible theoretically that another universes are more "METI-friendly" and they offer simpler solutions to achieve this kind of intergalactic messaging.

The highest level of the METI is the multiverse-METI. To realize it, it would be the first step to verify experimentally the existence of multiverses. E. g. one can search for neutrons which are leaking from a parallel braneworld. If one regards the universe as a system of "two braneworlds mutually invisible," then a neuron-shining-through-a-wall experiment would prove the existence of other braneworlds. [Sarrazin, Pignol et al 2015] Obviously it should be proved, too, that another braneworlds can be regarded as a "similar universes" and they are habitable for intelligent beings and even if the human race could find a technology to send artificial signs to another brane, it is problematic whether the message would be recognizable to the recipients. These are not new problems, since it is a question of the traditional METI whether there are other planets suitable for life and whether those intelligent inhabitants would able to decode the message.

Of course, one can play with the idea of creator who would be able to hide a message into the structure of a universe–just imagine that the numerical parts of every constant would be exactly 1. It is not evident that it is possible in the case of a physically existing sub universe, although in the case of a programmed world it should be feasible. On the other hand, these parameters can be interpreted by an intelligent habitant of that world as a coincidence.

This situation resembles both to the 17$^{th}$ century's debates about the presence or existence of the God of a clockwork universe and to the debate about our Universe's "fine-tuning" but despite of these uncertainties, multiverse hypothesis provides an opportunity to expand the scope of issues related to either extraterrestrial life, or SETI or METI, although it would be a logical error not to make a distinction between the phase-space of supposed possibilities and reality.

**References:**

- Adams, Fred and Laughlin, Greg: The Five Ages of the Universe. Inside the Physics of the Eternity. Touchstone Edition, 2000
- Aguirre, Anthony: The Cold Big-Bang Cosmology as a Counter-example to Several Anthropic Arguments, 2001 http://arxiv.org/abs/astro-ph/0106143
- Albrecht, Andreas – Sorbo, Lorenzo: Can the universe afford inflation? 2008 http://arxiv.org/pdf/hep-th/0405270v2.pdf
- Basalla, George: Civilized Life in the Universe: Scientists on Intelligent Extraterrestrials. Oxforf Univ. Press, 2006
- Barrow, John D. and Tipler, Frank J.: The Anthropic Cosmological Principle. Clarendon and Oxford Univ. Press, 1986
- Barrow, John D.: Pi in the Sky. Counting, Thinking and Being. Little, Brown and Company, 1992
- Bradbury, Robert J., Cirkovic, Milan M. and Dvorsky, George: Dysonian Approach to SETI: A Fruitful Middle Ground? JBIS, Vol. 64, pp.156-165, 2011
- Brooke, John Hedley: Science and Religion. Some Historical Perspectives. Cambridge Univ. Press 2014
- Carr, Bernard: Introduction and overview. In: Bernard Carr (editor): Universe or Multiverse? Cambridge Univ. Press, 2007
- Carrigan, Richard A. Jr.: Starry Messages: Searching for Signatures of Interstellar Archaeology, 2010. http://arxiv.org/abs/1001.5455
- Case-Winters, Anne: Reconstructing a Cristian Theology of nature. Ashgate Books, 2007
- Cirkovic, M. M., Dragicevic I and Beric-Bjedov, T.: ADAPTATIONISM FAILS TO RESOLVE FERMI'S PARADOX. Serb. Astron. J. 170 (2005), 89 – 100 http://saj.matf.bg.ac.rs/170/pdf/089-100.pdf
- cplire.ru (without author, without date): "Mir, Lenin, SSSR" http://www.cplire.ru/html/ra&sr/irm/MIR-LENIN-SSSR.html
- Cohen, Jack – Stewart, Ian: Evolving the Alien. The Science of Extraterrestrial Life. Ebury Press, 2002
- Collins, Robin: The Multiverse hypothesis: a theistic perspective. In: Bernard Carr (editor): Universe or Multiverse? Cambridge Univ. Press, 2007
- Fontenelle, Bernard de: Conversations on the Plurality of Worlds. Printed by J. Cundee, 1803. https://openlibrary.org/books/OL6981190M/Conversations_on_the_plurality_of_worlds
- Crowe, Michael J.: The Extraterrestrial Life Debate 1750-1900. The Idea of a Plurality of Worlds from Kant to Lowell. California Univ. Press, 1986
- Darling, David: Life Everywhere. The Maverick Science of Astrobiology. Basic Books, 2001

- Davies, Paul: The Last Three Minutes. Conjectures about the Ultimate Fate of the Universe. Basic Books, 1994
- Dick, Steven: The Biological Universe. The Twentieth-Century Extraterrestrial Life Debate and the Limits of Science. Cambridge Univ. Press, 1996
- Dick, Steven J.: The Postbiological Universe. 57th International Astronautical Congress, 2006 http://avsport.org/IAA/abst2006/IAC-06-A4.2.01.pdf
- Dick, Steven: The Twentieth Century History of the Extraterrestrial Life Debate: Major Themes and Lessons Learned. In: Vakoch, Douglas A. (editor): Astrobiology, History, and Society. Life Beyond Earth and the Impact of Discovery. Springer, 2013
- Dyson, Freeman: Time Without End: Physics and Biology in an Open Universe. Reviews of Modern Physics, Vol. 51, No. 3, July 1979
- Ellis, George: Multiverses: description, uniqueness and testing. In: Bernard Carr (editor): Universe or Multiverse? Cambridge Univ. Press, 2007
- Ellis, George F. R., Kirchner, U. and Stoeger, W. R.: Multiverses and Physical Cosmology. Mon.Not.Roy.Astron.Soc.347:921-936,2004
- Galantai, Zoltan: After Kardashev: A Farewell to Super Civilizations. Contact in Context, Vol 2. Issue 2., 2006
- Gale, George: Cosmological Fecundity: Theories of Multiple Universes. In: John Leslie (editor): Modern Cosmology and Philosophy. Prometheus Books, 1998
- Harrison, Albert A.: After Contact. The Human Response to Extraterrestrial Life. Perseus Publishing, 1997
- Harrison, Edward R.: Natural Selection of Universes Containing Intelligent Life. Q. J. R. astr. Soc. 1995, 36/193 – 203 http://articles.adsabs.harvard.edu/cgi-bin/nph-iarticle_query?bibcode=1995QJRAS..36..193H&db_key=AST&page_ind=0&data_type=GIF&type=SCREEN_VIEW&classic=YES
- Heller, René and Armstrong, John: Superhabitable Worlds. http://arxiv.org/abs/1401.2392, 2014
- Jenkins, Alessandro and Perez, Gilad: Looking for Life in the Multiverse Universes with different physical laws might still be habitable. Scientific American, Jan. 2010 http://www.scientificamerican.com/article/looking-for-life-in-the-multiverse/
- Junker, Thomas: Gesichte der Biologie. Die Wissenschraft wom Leben. Verlag C. H. Beck oHG, 2004
- Kardashev, N. S.: Transmission of Information by Extraterrestrial Civilizations. In: Soviet Astronomy-AJ Vol. 8. No. 2., sept. – oct. 1964 http://articles.adsabs.harvard.edu/cgi-bin/nph-iarticle_query?1964SvA.....8..217K&classic=YES
- Kawaler, S. and Veverka, J.: The Habitable Sun - One of Herschel, William's Stranger Ideas. Journal of the Royal Astronomical Society of Canada, Vol. 75, P. 46, 1981 http://adsabs.harvard.edu/abs/1981JRASC..75...46K
- Knoll, Andrew H. and Bambach, Richard K.: Directionality in the history of life: diffusion from the left wall or repeated scaling of the right? Paleobiology Vol. 26, No. 4, Autumn, 2000 http://isites.harvard.edu/fs/docs/icb.topic231281.files/Reading01_Lec24_OEB-113.pdf
- Lanier, Jaron: Rearranging Stars to Communicate with AliensA proposal to create special constellations that nature would never produce. Discover magazine (8 February 2008) http://discovermagazine.com/2008/feb/rearranging-stars-to-communicate-with-aliens
- Loeb, Avi: The Habitable Epoch of the Early Universe. 2014 http://arxiv.org/pdf/1312.0613v3

- Michaud, Michael A. G.: Contact with Alien Civilizations. Our Hopes and Fears about Encountering Extraterrestrials. Copernicus Books, 2007
- Rubinstein, Mary-Jane: Worlds without End. The Many Lives of the Universe. Columbia Univ. Press 2014.
- Sagan, Carl: The Cosmic Connection: An Extraterrestrial Perspective. Dell Publishing, 1975
- Sarrazin Michael; Pignol, Guillaume; Lamblin, Jacob; Petit, Fabrice; Terwagne, Guy; Nesvizhevsky, Valery V.: Probing braneworld hypothesis with a "neutron-shining-through-a-wall" experiment (15 January, 2015) http://arxiv.org/abs/1501.06468
- Shapiro, Robert and Feinberg, Gerald: Possible Forms of Life in Environments Very Different from the Earth. In: John Leslie (editor): Modern Cosmology and Philosophy. Prometheus Books, 1998
- Smolin, Lee: The status of cosmological natural selection (18 Dec 2006) http://arxiv.org/abs/hep-th/0612185
- Tegmark, Max: Parallel Universes. Scientific American (May 2003)
- Tsytovich, V N; G E Morfill, V E Fortov, N G Gusein-Zade, B A Klumov and S V Vladimirov; Fortov, V E; Gusein-Zade, N G; Klumov, B A; Vladimirov, S V: From plasma crystals and helical structurestowards inorganic living matter. New Journal of Physics (14 August 2007). http://www.iop.org/EJ/abstract/1367-2630/9/8/263

*I would like to say thanks to John D. Barrow for corrections about the historical origin of the natural selection theory of the universes and about the mathematical universe concept.*